\documentclass[conference]{IEEEtran}
\usepackage{graphicx} 
\DeclareGraphicsExtensions{.eps}

\usepackage{tikz}
\usepackage[dvipsnames]{xcolor}
\usepackage{pgfplots,pgfplotstable}
\usepackage{pgf-pie}
\pgfplotsset{compat=newest}
\usepgfplotslibrary{groupplots}
\usetikzlibrary{pgfplots.groupplots}
\pgfplotsset{compat=1.18}
\usepackage{siunitx}

\usepackage{multirow}
\usepackage{amsmath}
\usepackage{float}
\usepackage{upgreek}
\usepackage{booktabs}
\usepackage{enumitem}

\usepackage{svg}

\usepackage{textcomp}
\usetikzlibrary{shapes, arrows.meta, positioning}
\tikzset{
    block/.style = {rectangle, draw, align=center, minimum width=5.5cm, minimum height=1.2cm},
    arrow/.style = {thick, -{Stealth}}
}

\usepackage{empheq} 

\usepackage{stfloats} 

\usepackage[hyphens]{url}
\usepackage[hidelinks]{hyperref}
\hypersetup{colorlinks=true,linkcolor=black,citecolor=black,filecolor=black,urlcolor=black,breaklinks=true}
\usepackage{cite} 

\usepackage{cancel} 

\usepackage{array} 

\usepackage[export]{adjustbox} 

\usepackage[normalem]{ulem} 

\usepackage{amsfonts} 

\usepackage{amssymb} 

\makeatletter
\def\bstctlcite{\@ifnextchar[{\@bstctlcite}{\@bstctlcite[@auxout]}}
\def\@bstctlcite[#1]#2{\@bsphack
 \@for\@citeb:=#2\do{%
   \edef\@citeb{\expandafter\@firstofone\@citeb}%
   \if@filesw\immediate\write\csname #1\endcsname{\string\citation{\@citeb}}\fi}%
 \@esphack}
\makeatother

\begin{document}

\bstctlcite{IEEEexample:BSTcontrol}

\title{Evaluating the Impact of a Load Admittance Approximation in Transient Stability-Constrained Optimal Power Flow} 

\author{
\IEEEauthorblockN{Alex Junior da Cunha Coelho, Araceli Hernandez and Luis Badesa}
\vspace{1mm}
\IEEEauthorblockA{\textit{Technical University of Madrid (UPM), Madrid, Spain}
\\
\{alexjunior.dacunhacoelho, araceli.hernandez, luis.badesa\}@upm.es}
}



%


\maketitle

\begin{abstract}
The Transient Stability-Constrained Optimal Power Flow (TSC-OPF) incorporates dynamic stability constraints into the OPF formulation to ensure secure and economical operation under disturbances. While discretizing system dynamics enables the use of nonlinear programming techniques, it significantly increases computational burden. To enhance scalability, many studies simplify the network by representing loads as constant admittances, allowing the use of Kron reduction. However, computing the Kron reduction outside the optimization requires a voltage-based assumption to convert loads from constant power to constant admittance. This paper proposes a practical voltage-based load admittance approximation and evaluates the errors it may introduce in rotor angle and speed deviation trajectories. Case studies on the WECC 9-bus system show that the proposed approach reproduces rotor dynamics consistent with time-domain simulations during the first few seconds while considerably reducing implementation effort and mitigating convergence issues. The proposed framework thus offers a simple and effective strategy for scalable TSC-OPF implementations.
\end{abstract}


%
\IEEEpeerreviewmaketitle

\section*{Nomenclature}
\addcontentsline{toc}{section}{Nomenclature}

\subsection*{Indices and Sets}
\begin{IEEEdescription}[\IEEEusemathlabelsep\IEEEsetlabelwidth{$\lambda_1, \lambda_2, \lambda_3$}]
    \item[$k,m,\,\, \mathcal{N}$] Indices and set of buses
    \item[$g,i,\,\, \mathcal{G}$] Indices and set of generators
    \item[$d,\,\, \mathcal{D}$] Index and set of loads
    \item[$t,\,\, \mathcal{T}$] Index and set of time periods corresponding to the during-fault and post-fault stages
\end{IEEEdescription}

\subsection*{Constants and Parameters}
\begin{IEEEdescription}[\IEEEusemathlabelsep\IEEEsetlabelwidth{$\lambda_1, \lambda_2, \lambda_3$}]
    \item[$\mathrm{C}_g$] Cost function of $g$ (\texteuro$\mathrm{/MWh}$)
    \item[$\mathrm{P}_d, \mathrm{Q}_d$] Active and reactive power demanded by $d$ ($\mathrm{p.u.}$)
    \item[$\dot{\mathrm{Y}}, \mathrm{G}, \mathrm{B}$] Admittance matrix and its real and imaginary components ($\mathrm{p.u.}$)
    \item [$(\cdot)^{\mathrm{ref}}$] Denotes the slack bus in the pre-fault stage
    \item [$(\cdot)^{\mathrm{red}}$] Denotes the reduced version of the admittance matrix and its real and imaginary components
    \item[$\mathrm{\Delta t}$] Time-step of the dynamic simulation ($\mathrm{s}$)
    \item[$\upomega_{\mathrm{syn}}$] Synchronous angular frequency ($\mathrm{rad/s}$)
    \item[$\mathrm{x}^{\prime}_g$] Direct axis transient reactance of $g$ ($\mathrm{p.u.}$)
    \item[$\mathrm{H}_g$] Inertia constant of $g$ ($\mathrm{s}$)
    \item[$\mathrm{D}_g$] Damping coefficient of $g$ ($\mathrm{p.u.}$)

    \item [$\mathrm{P}_g^{\mathrm{min}}\, , \mathrm{P}_g^{\mathrm{max}}$] Min/Max active power generated by $g$ ($\mathrm{p.u.}$)
    \item [$\mathrm{Q}_g^{\mathrm{min}}, \mathrm{Q}_g^{\mathrm{max}}$] Min/Max reactive power generated by $g$ ($\mathrm{p.u.}$)
    \item [$\mathrm{V}_k^{\mathrm{min}}, \mathrm{V}_k^{\mathrm{max}}$] Min/Max voltage magnitude at bus $k$ ($\mathrm{p.u.}$)
    \item [$\uptheta_k^{\mathrm{min}} \, , \uptheta_k^{\mathrm{max}}$] Min/Max phase angle of the voltage at bus $k$ ($\mathrm{rad}$)
    \item [$\mathrm{S}_{km}^{\mathrm{max}}$] Maximum power transfer capacity of line ${km}$ ($\mathrm{p.u.}$)
    \item [$\uptheta_{km}^{\mathrm{min}} \, , \uptheta_{km}^{\mathrm{max}}$] Min/Max phase angle difference between adjacent buses $k$ and $m$ ($\mathrm{rad}$)
    \item [$\mathrm{E}_g^{\mathrm{min}}, \mathrm{E}_g^{\mathrm{max}}$] Min/Max internal voltage magnitude of $g$ ($\mathrm{p.u.}$)
    \item [$\updelta_{g}^{\mathrm{min}} \, ,\updelta_{g}^{\mathrm{max}}$] Min/Max rotor angle deviation of $g$ with respect to the center of inertia ($\mathrm{rad}$)
    
    
\end{IEEEdescription}

\subsection*{Decision Variables}
\begin{IEEEdescription}[\IEEEusemathlabelsep\IEEEsetlabelwidth{$\lambda_1, \lambda_2, \lambda_3$}]
    \item[$P_g, Q_g$] Active and reactive power produced by $g$ ($\mathrm{p.u.}$)
    \item[$V_k, \theta_{k}$] Magnitude and phase angle of the voltage at bus $k$ ($\mathrm{p.u.}, \, \mathrm{rad}$)
    \item[$P_{km}, Q_{km}$] Active and reactive power flow between buses $k$ and $m$ ($\mathrm{p.u.}$)
    \item[$E_g$] Magnitude of the internal voltage of $g$ ($\mathrm{p.u.}$)
    \item[$\delta_{g}^{0}$] Rotor angle of $g$ in the pre-fault ($\mathrm{rad}$)
    \item[$\Delta \omega_{g}^{0}$] Speed deviation of $g$ in the pre-fault ($\mathrm{p.u.}$)
    \item[$\delta_{g}^{t}$] Rotor angle of $g$ at time $t$ ($\mathrm{rad}$)
    \item[$\Delta \omega_{g}^{t}$] Speed deviation of the $g$ at time $t$ ($\mathrm{p.u.}$)
    \item[$P_{g}^{\mathrm{mec}}$] Mechanical power input of $g$ ($\mathrm{p.u.}$)
    \item[$P_{g}^{\mathrm{ele},t}$] Active electrical power output of $g$ at time $t$ ($\mathrm{p.u.}$)
    \item[$\delta_{COI}^{t}$] Rotor angle of the center of inertia at time $t$ ($\mathrm{rad}$)
    
      
\end{IEEEdescription}

\section{Introduction}
\label{sec:intro}
The Transient Stability-Constrained Optimal Power Flow (TSC-OPF) extends the conventional OPF formulation by explicitly incorporating dynamic stability constraints to ensure secure and economical system operation under credible disturbances \cite{Zimmerman2000}. These constraints are typically embedded by discretizing the differential-algebraic equations (DAEs) that govern system dynamics over a short time horizon, usually covering the first few seconds after fault inception \cite{Abhyankar2017, Liederer2022, Arredondo2023}. This discretization transforms the DAEs into a large set of algebraic equations solvable by standard nonlinear programming techniques \cite{Abhyankar2017}. However, it significantly increases problem size because each time step adds new decision variables and constraints to the problem, making the resulting optimization computationally intensive, especially for realistic network models and multiple contingency analyses \cite{Conejo2010}.

Due to scalability issues, some works have formulated the TSC-OPF using the classical (second-order) generator model, and have represented loads as constant power in the pre-fault stage and as constant admittances during the transient period after fault inception \cite{Conejo2010, Calle2013, Aghahassani2022}. The use of these models significantly reduce the computational effort required by optimization solvers, as the network can be simplified through Kron reduction \cite{Anderson2002}.

However, from an implementation standpoint, this simplified approach introduces a key challenge: converting loads to constant admittances for the during- and post-fault conditions requires the bus voltage magnitudes, which are unknown before solving the optimization. These admittances must also be incorporated into the network admittance matrix and reduced to the internal generator nodes via Kron reduction \cite{Anderson2002}. Embedding this reduction within the optimization framework is computationally demanding, as the augmented admittance matrix depends on decision variables, specifically the bus voltage magnitudes. Moreover, the Kron reduction requires inverting a complex-valued matrix, which may increase numerical complexity and hinder convergence.

The aforementioned literature lacks details on how loads are converted from constant power to constant admittances within the simplified TSC-OPF formulation. To address this gap, this paper explicitly proposes a practical approximation in which load admittances are obtained by dividing the nominal power by a voltage $1$ $\mathrm{p.u.}$ This simplification enables precomputing the reduced admittance matrix outside the optimization framework, thereby reducing implementation complexity. However, it may introduce inaccuracies in rotor angle and speed trajectories, particularly in AC-based TSC-OPF formulations where bus voltages typically deviate by $\pm \,5 \text{\textendash}10\%$ from their nominal value. To quantify the impact of this voltage assumption on load-admittance conversion, error metrics are employed.

The remainder of this paper is organized as follows. Section \ref{sec:methodology} introduces the TSC-OPF formulation using Kron reduction and discusses the voltage assumption on load-admittance conversion. Section \ref{sec:case_studies} presents case studies and analyzes the numerical results. Finally, Section \ref{sec:conclusion} summarizes the main conclusions and outlines directions for future work.

\section{Methodology}
\label{sec:methodology}
\subsection{TSC-OPF Formulation}
\label{subsec:swing}
Taking into account a single contingency in the stability assessment, here the AC-based TSC-OPF formulation with the classical generator model \cite{Anderson2002} is developed by representing loads as constant power in the steady-state solution (pre-fault) and as constant admittances during the dynamic period \cite{Conejo2010, Calle2013, Aghahassani2022}, which includes the during- and post-fault stages. The resulting mathematical formulation can be expressed as follows.
\begin{equation}
\hspace*{-38.5mm}\mathrm{min} \qquad f(P_{g}) = \sum_{g \, \in \, \mathcal{G}} \mathrm{C}_g \cdot P_{g}
\label{eq_obj}
\end{equation}
\hspace{2.0mm} subject to:
\begin{alignat}{1} 
    & \hspace*{3.3mm} P_{g,k} - \mathrm{P}_{d,k} = \sum_{m \, \in \, \mathcal{N}_k} P_{km} \label{eq_eqc_pbalance}\\
    & \hspace*{3.3mm} Q_{g,k} - \mathrm{Q}_{d,k} = \sum_{m \, \in \, \mathcal{N}_k} Q_{km} \label{eq_eqc_qbalance}\\
    & \hspace*{3.3mm} P_{km} = V_k \, V_m \, \left(\mathrm{G}_{km} \cos(\theta_{k} - \theta_{m}) + \mathrm{B}_{km} \sin(\theta_{k} - \theta_{m}) \right) \label{eq_eqc_pflow}\\
    & \hspace*{3.3mm} Q_{km} = V_k \, V_m \, \left(\mathrm{G}_{km} \sin(\theta_{k} - \theta_{m}) - \mathrm{B}_{km} \cos(\theta_{k} - \theta_{m}) \right) \label{eq_eqc_qflow}\\
    & \hspace*{3.3mm} \theta^{\mathrm{ref}} = 0 \label{eq_eqc_thetaref}\\
    & \hspace*{3.3mm} \mathrm{P}^{\mathrm{min}}_{g} \leq P_{g} \leq \mathrm{P}^{\mathrm{max}}_{g} \label{eq_ineqc_Pg}\\
    & \hspace*{3.3mm} \mathrm{Q}^{\mathrm{min}}_{g} \leq Q_{g} \leq \mathrm{Q}^{\mathrm{max}}_{g} \label{eq_ineqc_Qg}\\
    & \hspace*{3.3mm} \mathrm{V}^{\mathrm{min}}_{k} \leq V_{k} \leq \mathrm{V}^{\mathrm{max}}_{k} \label{eq_ineqc_voltage}\\
    & \hspace*{3.3mm} \uptheta^{\mathrm{min}}_{k} \leq \theta_{k} \leq \uptheta^{\mathrm{max}}_{k} \label{eq_ineqc_anglesnodes}\\
    & \hspace*{3.3mm} P^2_{km} + Q^2_{km} \leq \left(\mathrm{S}^{\mathrm{max}}_{km}\right)^2 \label{eq_ineqc_Siklim}\\
    & \hspace*{3.3mm} P^2_{mk} + Q^2_{mk} \leq \left(\mathrm{S}^{\mathrm{max}}_{km}\right)^2 \label{eq_ineqc_Skilim}\\
    & \hspace*{3.3mm} \uptheta^{\mathrm{min}}_{km} \leq \theta_{k} - \theta_{m} \leq \uptheta^{\mathrm{max}}_{km} \label{eq_ineqc_angledifferences}\\
    & \hspace*{3.3mm} P_{g,k} \mathrm{x}^{\prime}_g = E_g V_{k,g} \sin(\delta_g^0 - \theta_{k,g}) \label{eq_eqc_pinitial}\\
    & \hspace*{3.3mm} Q_{g,k} \mathrm{x}^{\prime}_g = - V_{k,g}^2 + E_g V_{k,g} \cos(\delta_g^0 - \theta_{k,g}) \label{eq_eqc_qinitial}\\
    & \hspace*{3.3mm} \Delta \omega_g^0 = 0 \label{eq_eqc_omegainitial}\\
    & \hspace*{3.3mm} \mathrm{E}^{\mathrm{min}}_{g} \leq E_{g} \leq \mathrm{E}^{\mathrm{max}}_{g} \label{eq_ineqc_voltagegen}\\
    & \hspace*{3.3mm} \delta_g^{t} - \delta_g^{t-1} =  \frac{\upomega_{\mathrm{syn}} \mathrm{\Delta t}}{2}  \left( \Delta \omega_g^{t} + \Delta \omega_g^{t-1} \right) \label{eq_eqc_delta}\\
    & \hspace*{3.3mm} \begin{aligned}
        & \Delta \omega_{g}^{t} \left(1 + \frac{\mathrm{D}_g \mathrm{\Delta t} }{4 \mathrm{H}_g}\right) - \Delta \omega_g^{t-1} \left(1 - \frac{\mathrm{D}_g\mathrm{\Delta t}}{4 \mathrm{H}_g}\right) \\ 
        & = \left(\frac{\mathrm{\Delta t}}{4 \mathrm{H}_g}\right) \left( 2P^{\mathrm{mec}}_{g} - P^{\mathrm{ele}, t}_{g} - P^{\mathrm{ele}, {t-1}}_{g} \right) \label{eq_eqc_omega}
    \end{aligned}\\
    & \hspace*{3.3mm} P^{\mathrm{ele}, t}_{g} = E_g \sum_{i \, \in \, \mathcal{G}} E_i \left( \mathrm{G}^{\mathrm{red}}_{g,i} \cos(\delta_g^t - \delta_i^t) + \mathrm{B}^{\mathrm{red}}_{g,i} \sin(\delta_g^t - \delta_i^t) \right) \label{eq_eqc_pele}\\
    & \hspace*{3.3mm} \delta_{\mathrm{COI}}^{t} = \frac{\sum_{g \, \in \mathcal{G}} \mathrm{H}_g \delta_g^t}{\sum_{i \, \in \mathcal{G}} \mathrm{H}_i} \label{eq_eqc_deltaCOI}\\
    & \hspace*{3.3mm} \updelta^{\min} \leq \delta_g^t - \delta_{\mathrm{COI}}^t \leq \updelta^{\max} \label{eq_ineqc_angle_COI}
\end{alignat} 
The objective function in \eqref{eq_obj} minimizes the total generation cost as a function of the active power output of each generator. Eqs. \eqref{eq_eqc_pbalance} and \eqref{eq_eqc_qbalance} represent the active and reactive power balance at each bus, respectively, during the pre-fault operating condition. Eqs. \eqref{eq_eqc_pflow} and \eqref{eq_eqc_qflow} define the active and reactive power flows through transmission lines and transformers connecting the buses. The voltage angle at the reference (slack) bus is fixed by \eqref{eq_eqc_thetaref}.

The operational constraints in \eqref{eq_ineqc_Pg}–\eqref{eq_ineqc_angledifferences} impose the system limits in the pre-fault stage. Specifically, \eqref{eq_ineqc_Pg} and \eqref{eq_ineqc_Qg} define the lower and upper bounds of the active and reactive power generation, respectively; \eqref{eq_ineqc_voltage} and \eqref{eq_ineqc_anglesnodes} constrain the bus voltage magnitudes and phase angles; \eqref{eq_ineqc_Siklim} and \eqref{eq_ineqc_Skilim} restrict the apparent power flows in each branch; and \eqref{eq_ineqc_angledifferences} limits the angle differences between adjacent buses.

Eqs. \eqref{eq_eqc_pinitial} and \eqref{eq_eqc_qinitial} compute the initial internal voltage magnitude $E_g$ and the rotor angle $\delta_g$ of each generator under steady-state conditions. The initial speed deviation is set to zero in \eqref{eq_eqc_omegainitial}, assuming all generators operate at synchronous speed prior to the disturbance. The bounds on the internal voltage magnitude of the generators are set by \eqref{eq_ineqc_voltagegen}.

Eqs. \eqref{eq_eqc_delta} and \eqref{eq_eqc_omega} represent the discretized form of the swing equation using the trapezoidal integration method \cite{Conejo2010, Liederer2022, Arredondo2023}. The electrical power output of each generator in the during- and post-fault intervals is determined by \eqref{eq_eqc_pele}, which is evaluated using the real and imaginary parts of the reduced admittance matrices obtained via Kron reduction for each operating stage. The center-of-inertia (COI) angular deviation at each time step is obtained from \eqref{eq_eqc_deltaCOI}. Finally, \eqref{eq_ineqc_angle_COI} constrains the rotor angle deviation of each generator with respect to the COI, following the formulations in \cite{Conejo2010, Geng2012, Liederer2022, Aghahassani2022, Arredondo2023}.

\subsection{Load Modelling and Kron Reduction}
\label{subsec:Kron_reduction}
The conversion of constant power loads into equivalent constant admittances at each bus is given by \eqref{eq_load_admittance}.
\begin{equation}
\dot{Y}_{d,k} = \frac{(P_{d,k} - j Q_{d,k})}{|V_k|^2}
\label{eq_load_admittance}
\end{equation}
The original bus admittance matrix $(\dot{\mathrm{Y}})$ is augmented to include the generator and load admittances. In this approach, only internal generator nodes inject current into the network, leading to the system relation in \eqref{eq_yaugmented} \cite{Anderson2002}.
\begin{equation}
\begin{bmatrix}
    \dot{Y}_{n_{\mathcal{N}} \times n_{\mathcal{N}}} & \dot{Y}_{n_{\mathcal{N}} \times n_{\mathcal{G}}} \\
    \dot{Y}_{n_{\mathcal{G}} \times n_{\mathcal{N}}} & \dot{Y}_{n_{\mathcal{G}} \times n_{\mathcal{G}}}
\end{bmatrix}
\cdot 
\begin{bmatrix}
    \dot{V}_{n_{\mathcal{N}}} \\
    \dot{E}_{n_{\mathcal{G}}}
\end{bmatrix}
    =
\begin{bmatrix}
    \dot{0}_{n_{\mathcal{N}}} \\
    \dot{I}_{n_{\mathcal{G}}}
\end{bmatrix}
\label{eq_yaugmented}
\end{equation}
where ${n_{\mathcal{N}}}$ and ${n_{\mathcal{G}}}$ are the number of buses and generators in the system, respectively. Here, ${\dot{V}}$ and ${\dot{E}}$ denote the complex voltages in the buses and the internal generator nodes, respectively. Vector ${\dot{I}}$ is the injected generator currents, and the submatrices of ${\dot{Y}}$ represent the corresponding network couplings. Matrix ${\dot{Y}}_{n_{\mathcal{N}} \times n_{\mathcal{N}}}$ is similar to the bus admittance matrix, except that the generator and load admittances are included in it; while matrix ${\dot{Y}}_{n_{\mathcal{G}} \times n_{\mathcal{G}}}$ is a diagonal matrix of generator admittances. Matrix ${\dot{Y}}_{n_{\mathcal{N}} \times n_{\mathcal{G}}}$ is the negative of the generator admittance at the position of the internal generator node, and zero otherwise. Finally, matrix ${\dot{Y}}_{n_{\mathcal{G}} \times n_{\mathcal{N}}}$ is the transpose of ${\dot{Y}}_{n_{\mathcal{N}} \times n_{\mathcal{G}}}$ \cite{Anderson2002}.

Since only internal generator nodes inject current, the admittance matrix can be reduced by applying \eqref{eq_yred}.
\begin{equation}
    {\dot{Y}}^{\mathrm{red}} = {\dot{Y}}_{n_{\mathcal{G}} \times n_{\mathcal{G}}} - \left[ {\dot{Y}}_{n_{\mathcal{G}} \times n_{\mathcal{N}}} \left( {\dot{Y}}^{-1}_{n_{\mathcal{N}} \times n_{\mathcal{N}}} \right)  {\dot{Y}}_{n_{\mathcal{N}} \times n_{\mathcal{G}}}\right]
    \label{eq_yred}
\end{equation}

When constructing the TSC-OPF model, different reduced admittance matrices are defined for the during- and post-fault periods, so that the condition of the network is properly reflected in the reduced equivalent. For instance, a three-phase bolted short circuit at bus $k$ can be represented by adding a large shunt admittance (e.g., $10^6$ p.u.) to the diagonal element $(k,k)$ of the bus admittance matrix ${Y}_{n_{\mathcal{N}} \times n_{\mathcal{N}}}$. Similarly, the disconnection of a transmission line between buses $k$ and $m$ is modeled by removing the corresponding branch admittances from the matrix.

As discussed in Section \ref{sec:intro}, embedding the Kron reduction within the optimization framework and updating it at every solver iteration can be computationally burdensome and may hinder convergence. To address this issue, this paper proposes a simple yet physically consistent strategy: loads are converted from constant power to constant admittance prior to building the optimization model, assuming a voltage magnitude of $1.0 \, \mathrm{p.u.}$ at the load buses. These admittances are then incorporated into the Kron reduction for both the during- and post-fault stages, yielding reduced network models that are precomputed and subsequently used in the optimization formulation. It is important to note that, during the optimization, the voltage magnitudes at all buses are still allowed to vary within the limits specified by \eqref{eq_ineqc_voltage}.

\subsection{Comparison Procedure}
\label{subsec:comparison_methodology}
To evaluate the impact of the proposed voltage assumption for load–admittance conversion, the results obtained from the TSC-OPF are compared with benchmark results from a time-domain software, using the following steps:
\begin{enumerate}[label=\textbf{Step~\arabic* --}, leftmargin=1.48cm]
\item Solve an AC-OPF for the network without transient stability constraints.
\item For a given contingency, run the TSC-OPF using the load–admittance approximation. This case is referred to as without (w/o) correction.
\item Repeat Step 2, but update the load admittances according to the bus voltage magnitudes obtained in Step 2. This case is referred to as with (w) correction.
\item Using a benchmark transient stability tool, simulate the same contingency considering the bus voltages and generator dispatch from Step 3.
\item Compare the results from Steps 2 and 3 with the benchmark results from Step 4.
\end{enumerate}

For each contingency, Steps 2–5 are performed using two time steps, $\mathrm{\Delta t} = 10 \,\mathrm{ms}$ and $\mathrm{\Delta t} = 1 \,\mathrm{ms}$, to evaluate the effect of temporal resolution on solution accuracy. Benchmark simulations are carried out using a time step of $\mathrm{\Delta t} = 1 \,\mathrm{ms}$ with the academic version of the ANATEM time-domain simulator developed by CEPEL~\cite{CEPEL_ANATEM}. ANATEM was selected because it also employs the trapezoidal method to solve the DAEs. The mean errors between results obtained with different time steps are computed after interpolating the trajectories.

\section{Case Studies}
\label{sec:case_studies}
\begin{figure}[tb!]
    \centering
    \includegraphics[width=0.75\columnwidth, height=0.35\columnwidth]{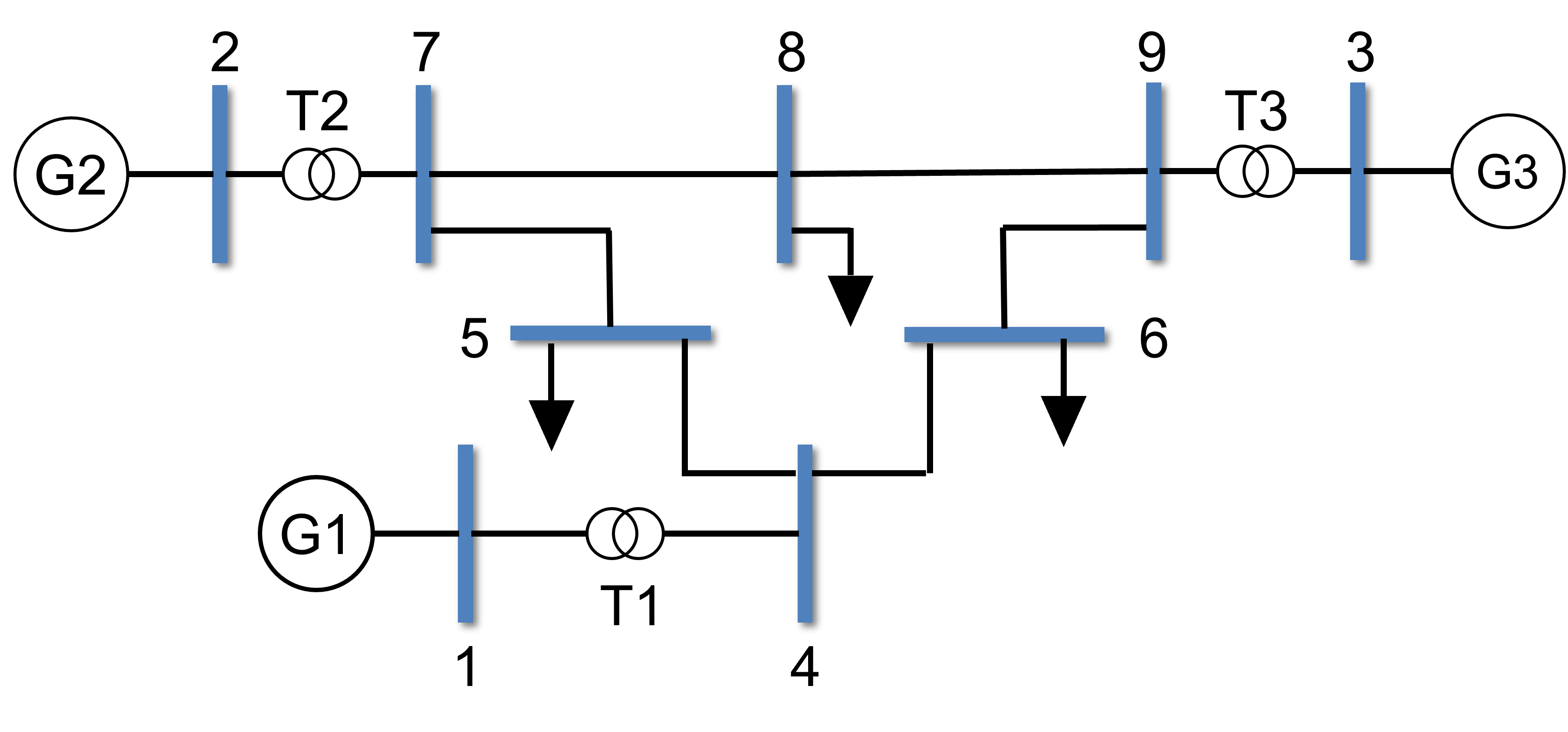}
    \caption{Single-line diagram of the WECC 9-bus, 3-machine system.}
    \label{fig:wecc9b3m}
\end{figure}

In this section, case studies are performed on the WECC 9-bus, 3-machine system, whose topology is shown in Fig.~\ref{fig:wecc9b3m}. System data and generation costs are from the MATPOWER database~\cite{MATPOWER}, and dynamic parameters from~\cite{pavella2000book}. To represent a more stressed operating condition, the active and reactive power of all loads in the original system are increased by a factor of $1.5$. The optimization model is implemented in \texttt{Julia-JuMP} \cite{JuMP2023} and solved using \texttt{IPOPT 3.14}. A simulation time window of $5 \, \mathrm{s}$ is considered, which is sufficient to capture the stability response during the first swings. The discretization of the swing equation is applied throughout the simulation period. The stability limits in \eqref{eq_ineqc_angle_COI} are set to $\pm 100^{\circ}$ w.r.t. COI, following \cite{Conejo2010, Geng2012, Liederer2022}. All code and data are publicly available at the repository \cite{Code}. 

\subsection{Contingency 1}
Contingency 1 consists of a three-phase short circuit at bus 4 occurring at $t = 0$ and cleared by opening line 4–5. The fault clearing time is set to $150 \, \mathrm{ms}$, which, in this case, is short enough to prevent \eqref{eq_ineqc_angle_COI} from becoming active. Consequently, the contingency does not alter the dispatch determined in Step~1.

For all simulations performed in Steps 1–3 of the proposed procedure, the resulting bus voltages are $V = [1.1;\,\allowbreak 1.1;\,\allowbreak 1.1;\,\allowbreak 1.0736;\,\allowbreak \mathbf{1.0294};\,\allowbreak \mathbf{1.0527};\,\allowbreak 1.0857;\,\allowbreak \mathbf{1.0688};\,\allowbreak 1.0957]\,\allowbreak\mathrm{p.u.}$, and generation dispatch values are $P_g = [1.4308;\,\allowbreak 1.9825;\,\allowbreak 1.3891]\,\allowbreak\mathrm{p.u.}$ and $Q_g = [0.5532;\,\allowbreak 0.3552;\,\allowbreak 0.1274]\,\allowbreak\mathrm{p.u.}$ These results show that the voltage magnitudes at the load buses (5, 6, and 8) are greater than $1.0 \, \mathrm{p.u.}$, suggesting that the rotor angle and speed deviation trajectories may differ from the benchmark in the cases w/o correction. This effect is quantified by computing the Mean Absolute Error (MAE) between the trajectories obtained from the TSC-OPF and those from the benchmark simulations, as summarized in Table~\ref{tab:errors_case_1}. Table~\ref{tab:errors_case_1} also reports the MAE for ANATEM simulated with $\mathrm{\Delta t} = 10 \,\mathrm{ms}$.
\begin{table}[!tb]
\caption{MAEs Obtained for Contingency 1}
\label{tab:errors_case_1}
\centering
\resizebox{0.95\columnwidth}{!}{
\begin{tabular}{ccccccc} \cline{3-7}
    & & \multicolumn{2}{c}{TSC-OPF} & \multicolumn{2}{c}{TSC-OPF} & \multirow{2}{*}{ANATEM} \\
    & & \multicolumn{2}{c}{w/o correction} & \multicolumn{2}{c}{w correction} & \\
    & & $10 \, \mathrm{ms}$ & $1 \, \mathrm{ms}$ & $10 \, \mathrm{ms}$ & $1 \, \mathrm{ms}$ & $10 \, \mathrm{ms}$ \\\hline
    \multirow{3}{*}{$\delta$}  & G1 & 1.5365 &  0.9896 & 0.5728 & 0.0611 & 0.2912\\ 
     & G2 & 4.7416 & 3.0724 & 1.7792 & 0.1866 & 0.8747\\
     & G3 & 4.8116 & 3.2151 & 1.1840 & 0.1049 & 0.4424 \\ \hline
    \multirow{3}{*}{$\Delta \omega$}  & G1 & 0.0071 & 0.0082 & 0.0013 & 0.0001343 & 0.0004831\\ 
     & G2 & 0.0073 & 0.0083 & 0.0013 & 0.0001311 & 0.0004849\\
     & G3 & 0.0071 & 0.0082 & 0.0013 & 0.0001316 & 0.0004760\\
    
    \hline
\end{tabular}
}
\end{table}

\begin{figure}[tb!]
    \centering
    \includegraphics[width=0.99\columnwidth]{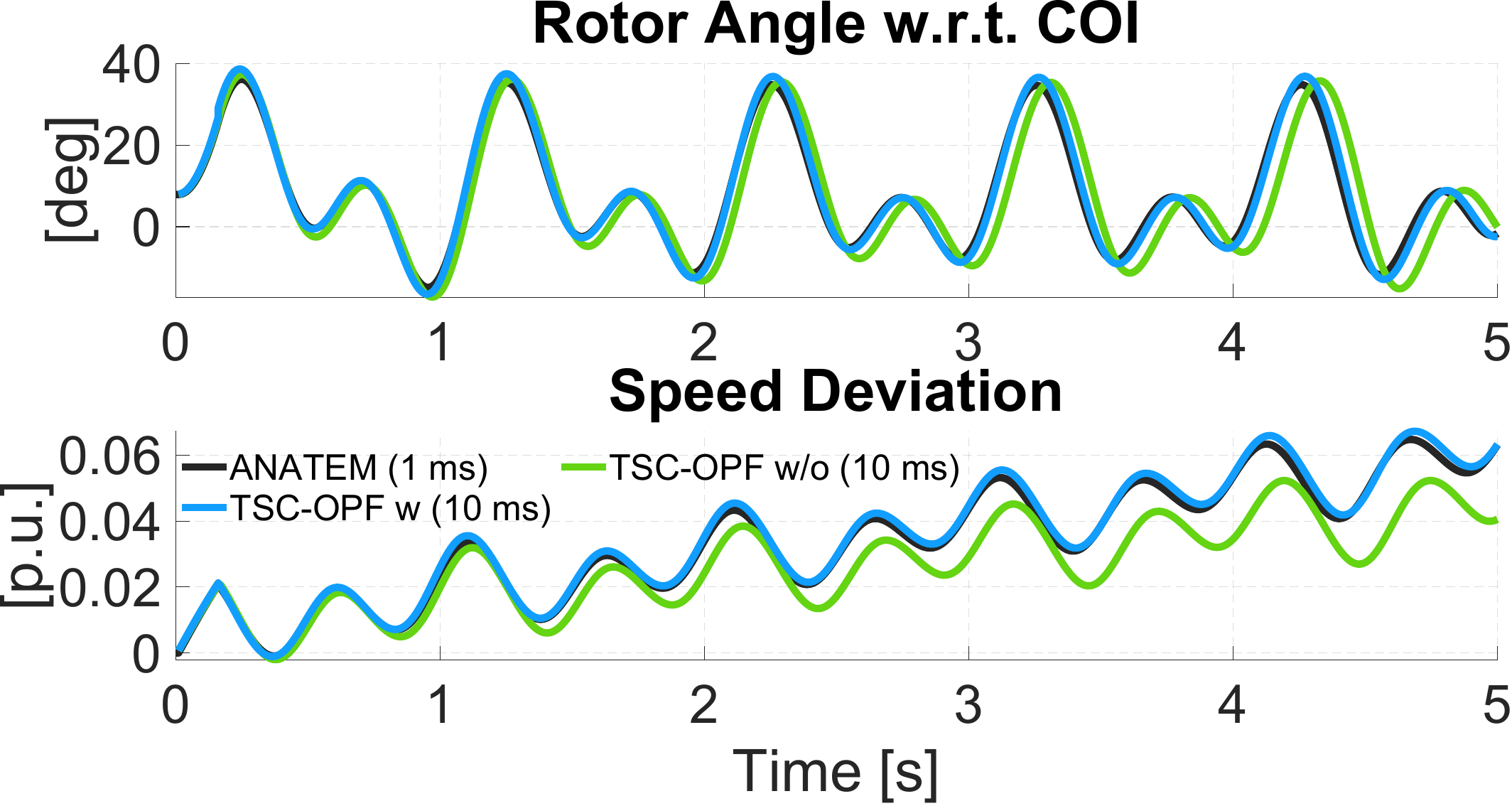}
    \caption{Contingency 1 - Comparison of rotor angle and speed deviation trajectories for G3 obtained with TSC-OPF w/o and w correction using $\mathrm{\Delta t} = 10 \,\mathrm{ms}$ against the benchmark (i.e., ANATEM with $\mathrm{\Delta t} = 1 \,\mathrm{ms}$).}
    \label{fig:gen2_10ms}
\end{figure}

Table~\ref{tab:errors_case_1} shows that the case w/o correction yields the highest MAE values for both analyzed time steps. As expected, the MAEs decrease when the load admittances are computed using the actual bus voltages obtained from the optimization, and also when the time step is reduced. Among the generators, G3, which has the lowest inertia constant, exhibits the largest MAEs for $\delta$ in the case w/o correction. The corresponding rotor angle w.r.t COI and speed deviation trajectories for G3, obtained with TSC-OPF using $\mathrm{\Delta t} = 10 \,\mathrm{ms}$ are plotted against the results obtained with ANATEM using $\mathrm{\Delta t} = 1 \,\mathrm{ms}$, as shown in Fig.~\ref{fig:gen2_10ms}.

Fig.~\ref{fig:gen2_10ms} shows that, despite the larger MAEs, the approximate approach still reproduces the main dynamic behavior of the rotor angle during the first swings, even with a relatively large time step. The rotor angle trajectories remain close to the benchmark, particularly in magnitude, whereas the speed deviations exhibit more noticeable differences that become significant only after approximately $t = 2 \, \mathrm{s}$. It is worth noting that several studies in the literature typically simulate the TSC-OPF over short time horizons, up to $2 \, \mathrm{s}$ ~\cite{Conejo2010, Calle2013, Aghahassani2022}. Under such conditions, the proposed voltage assumption has a limited impact on the results. Furthermore, the effect of the admittance approximation is more pronounced in the post-fault period. During the fault, the reduced admittance matrix is dominated by the high conductance associated with the three-phase short-circuit, so the voltage assumption has little influence on the results at this stage. 

\subsection{Contingency 2}
Contingency 2 represents a more stressed scenario from the dynamic perspective. It consists of a three-phase short circuit at bus 7 that occurs at $t = 0$ and is cleared by opening line 7–5. The fault clearing time is set to $300\,\mathrm{ms}$, so that the stability constraints force the optimization solver to identify a new dispatch different from the one obtained in Step~1 of the comparison procedure.

For all simulations performed in Steps 2 and 3 of the proposed procedure, the resulting bus voltages are $V = [1.1;\,\allowbreak 1.1;\,\allowbreak 1.1;\,\allowbreak 1.0755;\,\allowbreak \mathbf{1.0343};\,\allowbreak \mathbf{1.0555};\,\allowbreak 1.0868;\,\allowbreak \mathbf{1.0694};\,\allowbreak 1.0958]\,\allowbreak\mathrm{p.u.}$, and the generation dispatch values are $P_g = [2.2131;\,\allowbreak 1.2625;\,\allowbreak 1.3079]\,\allowbreak\mathrm{p.u.}$ and $Q_g = [0.5868;\,\allowbreak 0.2736;\,\allowbreak 0.121]\,\allowbreak\mathrm{p.u.}$ Compared to Contingency 1, the voltage magnitudes at the load buses increased slightly, whereas the active power dispatch changed significantly. The MAEs for the angle and speed deviation trajectories for this contingency are presented in Table~\ref{tab:errors_case_2}.
\begin{table}[!tb]
\caption{MAEs Obtained for Contingency 2}
\label{tab:errors_case_2}
\centering
\resizebox{0.95\columnwidth}{!}{
\begin{tabular}{ccccccc} \cline{3-7}
    & & \multicolumn{2}{c}{TSC-OPF} & \multicolumn{2}{c}{TSC-OPF} & \multirow{2}{*}{ANATEM} \\
    & & \multicolumn{2}{c}{w/o correction} & \multicolumn{2}{c}{w correction} & \\
    & & $10 \, \mathrm{ms}$ & $1 \, \mathrm{ms}$ & $10 \, \mathrm{ms}$ & $1 \, \mathrm{ms}$ & $10 \, \mathrm{ms}$ \\\hline
    
    \multirow{3}{*}{$\delta$}  & G1 & 5.4473 & 4.8983 & 2.7684 & 2.9088 & 5.2129\\ 
     & G2 & 16.3495 & 14.7099 & 8.1949 & 8.6702 & 15.5314\\
     & G3 & 14.2948 & 12.6282 & 7.1944 & 7.1819 & 11.9366 \\ \hline
     
    \multirow{3}{*}{$\Delta \omega$}  & G1 & 0.0059 & 0.0068 & 0.0022 & 0.0014 & 0.0029\\ 
     & G2 & 0.0069 & 0.0074 & 0.0035 & 0.0032 & 0.0061\\
     & G3 & 0.0093 & 0.0092 & 0.0041 & 0.0037 & 0.0062\\
    
    \hline
\end{tabular}
}
\end{table}

\begin{figure}[tb!]
    \centering
    \includegraphics[width=0.99\columnwidth]{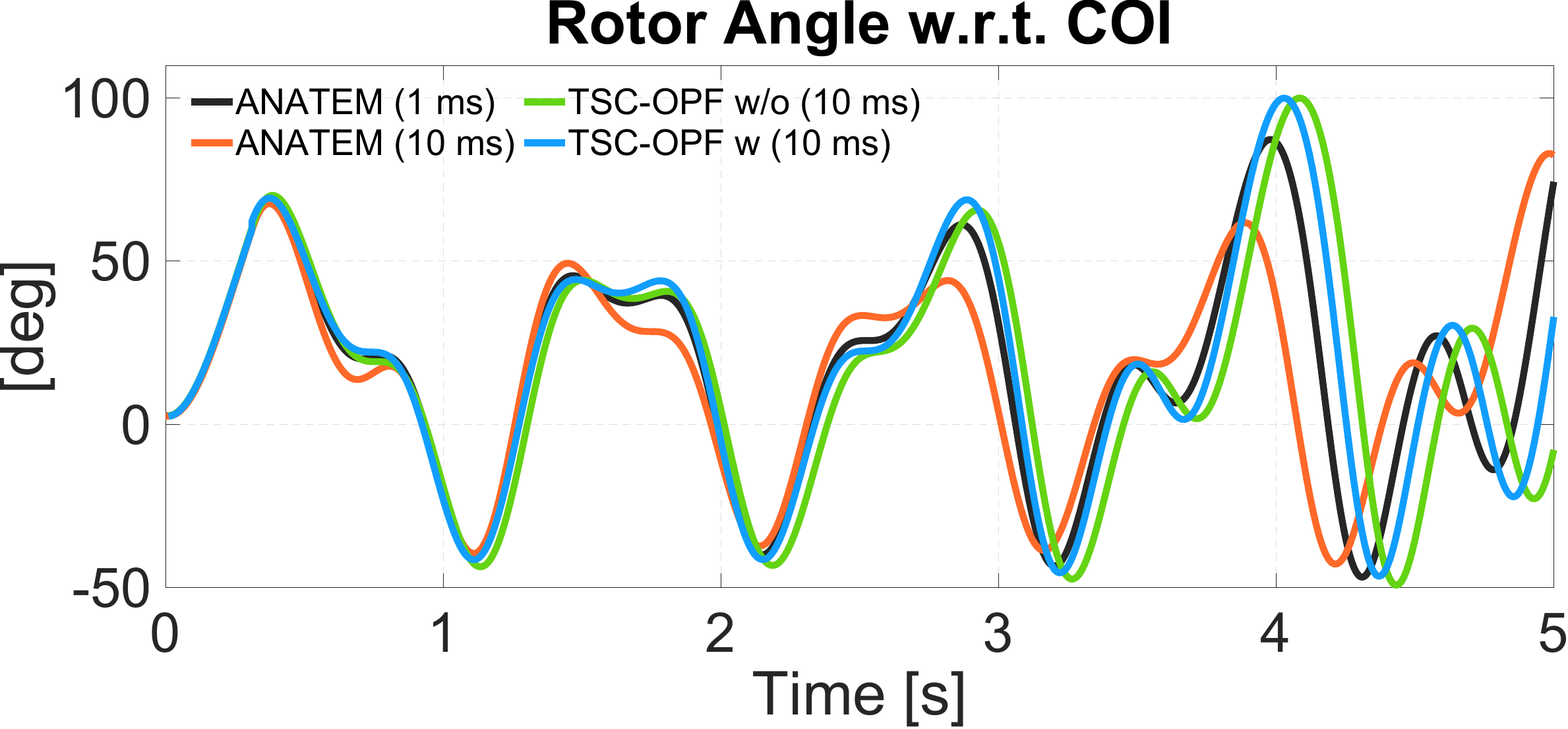}
    \caption{Contingency 2 - Comparison of rotor angle trajectories for G3 obtained with TSC-OPF w/o and w correction and ANATEM using $\mathrm{\Delta t} = 10 \,\mathrm{ms}$ against the benchmark (i.e., ANATEM with $\mathrm{\Delta t} = 1 \,\mathrm{ms}$).}
    \label{fig:gen3_contingency2}
\end{figure}

The results in Table~\ref{tab:errors_case_2} indicate that, under this more stressed condition, the voltage assumption leads to higher MAEs than those observed in Contingency 1, even for smaller time steps. It is noteworthy that the MAES obtained with ANATEM using $\mathrm{\Delta t} = 10\,\mathrm{ms}$ are non-negligible, being comparable with those from the TSC-OPF w/o correction. Once again, the TSC-OPF with correction produces trajectories that more closely match the benchmark, and the differences between the two time steps are less pronounced than in Contingency~1. Although G2 exhibits the highest MAE for $\delta$, G3 is the only generator that activates the stability constraint within the simulation time window. For comparison and analysis, the rotor angle trajectories of G3 w.r.t. COI obtained with TSC-OPF and ANATEM are shown in Fig.~\ref{fig:gen3_contingency2}.

From the curves in Fig.~\ref{fig:gen3_contingency2}, it can be seen that up to $t = 3 \, \mathrm{s}$, the trajectory obtained with the TSC-OPF w/o correction closely follows the benchmark, even with a time step of $\mathrm{\Delta t} = 10 \,\mathrm{ms}$, a feature that is crucial for scaling the problem to larger networks. Overall, the TSC-OPF results capture a similar dynamic trend during the first swings. More importantly, the trajectory produced using the proposed voltage assumption does not diverge significantly in amplitude, which is crucial to maintaining transient stability. Furthermore, it is evident that under more severe disturbances, even time-domain simulations can exhibit noticeable deviations when different integration time steps are employed.

\section{Conclusion}
\label{sec:conclusion}
This paper investigated the effect of converting loads from constant power to constant admittances by assuming a voltage magnitude of $1 \, \mathrm{p.u.}$ when constructing the reduced network model for the TSC-OPF formulation. Case studies on the WECC 9-bus system showed that, for both mild and severe contingencies, the voltage-based admittance assumption produces rotor angle trajectories that closely follow those from dedicated time-domain software, even with larger integration time steps. Although minor MAE differences were observed, the proposed approach effectively captures the dynamic behavior of the system while significantly reducing the implementation effort and mitigating potential convergence issues that could arise if the Kron reduction were performed within the optimization process. Hence, adopting the proposed framework represents a practical strategy for TSC-OPF implementations. Future work will further assess and compare the performance of the full and reduced admittance matrix formulations.

\section{Acknowledgements}
This work was supported by MICIU/AEI/10.13039/501100011033 and ERDF/EU under grants PID2023-150401OA-C22 and PID2022-141609OB-I00, as well as by the Madrid Government (Comunidad de Madrid-Spain) under the Multiannual Agreement 2023-2026 with Universidad Politécnica de Madrid, `Line A - Emerging PIs' (grant number: 24-DWGG5L-33-SMHGZ1).

\bibliographystyle{IEEEtran} 
\bibliography{Bibliography}

\end{document}